\documentclass[twocolumn,showpacs,preprintnumbers,amsmath,amssymb,
prl]{revtex4}
\usepackage{graphicx}
\usepackage{dcolumn}
\usepackage{bm}
\begin{document}

\title{Anomalous Negative Magnetoresistance Caused by
Non-Markovian Effects}

\author{Vadim~V.~Cheianov$^{1}$, A.~P.~Dmitriev
$^{2}$, V.~Yu.~Kachorovskii$^{2}$} \affiliation{$^1$NORDITA,
Blegdamsvej 17, Copenhagen,  DK 2100, Denmark
\\ $^2$A.F.~Ioffe Physical-Technical Institute, 26
Polytechnicheskaya str., Saint Petersburg, 194021, Russia}

\date{\today}

\begin{abstract} A theory of recently discovered
anomalous low-field magnetoresistance is developed
for the system of two-dimensional electrons scattered
by hard disks of radius $a,$ randomly distributed
with concentration $n.$ For small magnetic fields the
magentoresistance is found to be parabolic and
inversely proportional to the gas parameter, $ \delta
\rho_{xx}/\rho \sim -
(\omega_c \tau)^2 / n a^2.$ With increasing field the
magnetoresistance becomes linear $\delta
\rho_{xx}/\rho \sim  - \omega_c \tau $ in a good agreement with
the experiment and numerical simulations.

\end{abstract}


\maketitle

It is well known that in the Boltzmann-Drude approach the
longitudinal resistivity $\rho_{xx}$ of a degenerate
two-dimensional (2D) electron gas does not depend on the
transverse magnetic field $B$. Therefore, the known mechanisms of
magnetoresistance (MR) involve either quantum interference effects
or classical non-Markovian memory effects, which are not captured
in the Boltzmann picture.  The MR, arising from quantum effects
was discussed in a great number of works (see for review
Ref.~\cite{lee}). The role of classical memory effects was
underappreciated for a long time, though several theoretical works
pointed out at the importance of such effects for magnetotransport
\cite{baskin,polyakov,boby}.  The interest to the problem of
classical MR has sharply increased in recent years, starting with
Ref.~\cite{perel}, where it was shown that effects of "classical
localization" may lead to the exponential suppression of electron
diffusion at large $B$. This work was followed by a series of
works \cite{fog,bobynew,basknew,kuzm,mir1,mir2,mir3,igor,dmit},
discussing different aspects of classical 2D magnetotransport in
strong magnetic fields.

In this paper we focus on a mechanism of low field classical MR
specific for systems of strong scatterers. This mechanism is
connected to the memory effects due to backscattering events. The
corresponding corrections to the conductivity are small in the
parameter $a/l,$ where $a$ is the characteristic size of the
scatterers and $l$ is the mean free path. Nevertheless, the
dependance of these corrections on the magnetic field turns out to
be very sharp, resulting in the MR anomaly. The anomaly was
discovered in recent numerical simulations \cite{dmit1} where the
MR in a system of 2D electrons scattering on randomly distributed
hard disks was studied. This system is usually referred to as the
Lorenz gas and is the simplest model of the 2D electron gas with
strong scatterers. In the following we restrict our considerations
to this model. The generalization of our results to other models
of strong disorder is straightforward. Magnetotransport in the
Lorenz gas is characterized by two dimensionless parameters: $
\beta= \omega_c \tau ,$ and the gas parameter $\beta_0=a/l=2 n a^2
.$ Here $a$ is the disk radius, $n$ is disks' concentration,
$\omega_c$ is the cyclotron frequency, $\tau=l/v_F$ is the mean
free time and $l=1/2 n a$ is the mean free path. The anomaly was
observed in the case $\beta \ll 1 , \quad \beta_0 \ll 1.$ Both the
numerical simulations and the qualitative analysis of \cite{dmit1}
indicated that at zero temperature, $T,$ the MR can be expressed
in terms of a dimensionless function $f(z)$ via
\begin{equation} \frac{\delta\rho_{xx}}{\rho}
=-\beta_0 f\left(\frac{\beta}{\beta_0}\right),
\label{1}
\end{equation}
where $\rho $ is the resistivity for $B=0.$ Numerical results
\cite{dmit1} suggest that $f(z)\sim z$ as $z\to 0,$ yielding
\mbox{$\delta\rho_{xx}/\rho \sim -|\omega_c|\tau .$} The latter
expression is in a very good agreement with experimental
measurements of MR in a random antidot arrays \cite{dot1}. It is
anomalous in two senses.  First, it has a non-analytic dependence
on the magnetic field. Second, it does not vanish in the limit of
vanishing $\beta_0$, which is normally regarded as the expansion
parameter for the corrections to the Drude-Boltzmann picture. This
intriguing behavior calls for a rigorous analytical theory of the
effect, which would establish Eq. ~\ref{1} and enable one to
derive the analytical expression for function $f$. In this letter
we present a theory of the anomaly and give an expression for
$f(z).$ We find that for $z\lesssim 1,$ $f(z)$ is linear in
agreement with numerical experiment, but at very small $z\lesssim
0.05$ crosses over to a quadratic dependence.  Thus, for $\beta\to
0,$ Eq. ~\ref{1} yields $\delta \rho_{xx}/\rho \sim
-\beta^2/\beta_0.$ The limit $\beta_0\to 0$ should be taken with
care. While the small $\beta$ expansion seems to be singular as a
function of $\beta_0,$ the region of $\beta$ where this expansion
is valid shrinks as $\beta_0\to 0.$ For $z \to \infty,$ $f$
saturates at some constant value. Therefore, the full variation of
$\delta \rho_{xx}/\rho$ is of the order $\beta_0.$ In other words,
the anomalous MR is strong but it exists in a small region of
magnetic fields.

In \cite{dmit1} a mechanism of MR connected with memory effects
arising in backscattering events was proposed. It has a close
relation to the well known non-analyticity of the virial expansion
of transport coefficients \cite{dorfman,wei,wei1,hauge,peierls},
which we briefly recall. For $B=0$ the leading nonanalytic
correction to resistivity, $\delta \rho$, is due to the processes
of return to a scatterer after a single collision with another
scatterer [Fig.~1(a)].
\begin{figure}
\includegraphics[width=0.4\textwidth]{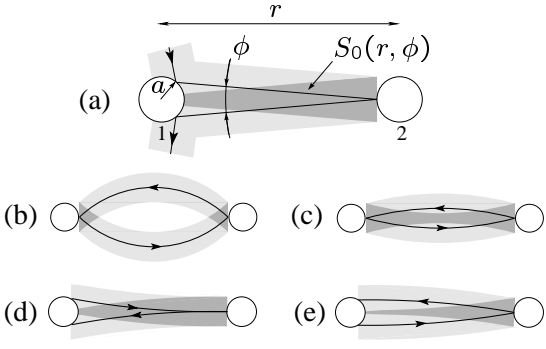}
\caption{ Backscattering process responsible for leading
nonanalytic contribution to the resistivity at $B=0$ (a).  For $B
\neq 0,$ the overlap area, $S_B,$ between two corridors is small
at large $B$ (b). For $\phi=0,$ $S_B$ decreases with $B$ (c). For
$\phi \neq 0$ and small $B$ the values of $S_B- S_0$ for time
reversed trajectories have opposite signs (d,e).} \label{fig1}
\end{figure}
The relative correction, $\delta \rho/\rho,$ is proportional to
the corresponding backscattering probability, given by the product
of $ e^ {-r/l} d \phi dr/l $ (which is the probability to reach
scatterer 2 without collision and scatter in the angle $d\phi$)
and the probability $p$ to return without collisions from 2 to 1
(here $l$ is the mean free path).  Assuming $p=\exp(-r/l)$ and
integrating over intervals $ 0< \phi< a/r, \quad a<r<\infty,$ one
obtains \cite{dorfman,wei,wei1, hauge,peierls} $\delta \rho /\rho
\sim \beta_0 \ln(1/2\beta_0).$

In Ref.~\cite{dmit1} it was shown that the
probability $p$ is actually larger than $\exp(-r/l)$
because the passage of a particle from $1$ to $2$
ensures the existence of a corridor of width $2a$
free of the centers of the disks. This reduces the
scattering probability on the way back, yielding
$p(r,\phi)=\exp(-r/l +n S_0(r,\phi)),$ where $ S_0(r,
\phi) = 2 a r -r^2
|\phi|/2 $ is the area of the overlap of the two
corridors [Fig.~1(a)].  For example, for $\phi=0,$ we
have $S_0=2ar$ and $p=1.$ Physically, this means that
the particle is unable to scatter, since it travels
back along the same path.  Taking into account the
effect of ``empty corridor'', we get
\begin{equation} \frac{\delta \rho}{\rho} \sim
\int_a^{\infty} \frac{dr}{l}
\int_0^{a/r} d\phi \ e^{-(2r/l) +nS_0} \approx
\beta_0 \ln \left(\frac{C}
{2 \beta_0}\right), \label{estimate} \end{equation}
where $C$ is a constant of the order of unity.  Thus,
for $B=0$ this effect simply changes the constant in
the argument of the logarithm.

The key idea suggested in \cite{dmit1} was that for
$B\neq 0$ the area of the overlap of the two
corridors, $S_B,$ sharply depends on $B,$ resulting
in the observed MR.  Indeed, it is seen from
Fig.~1(b) that for $\beta \gtrsim \beta_0 ,$ $S_B \to
0,$ resulting in sharp negative MR
\begin{equation} \frac{\delta \rho_{xx}}{\rho} \sim
\int_0^{\infty} \frac{dr}{l}
\int_0^{a/r} d\phi \ e^{-2r/l} \left ( e^{n S_B}
-e^{n S_0} \right).  \label{estimateB} \end{equation}
The following qualitative explanation of the observed
linear MR was presented in Ref. \cite{dmit1}.  The
value $n ( S_B - S_0)$ was estimated for $\phi=0$
[Fig.~1(c)] to the first order in $B$ as $- nr^3/R_c
=- r^3/2 a l R_c,$ where $R_c$ is the cyclotron
radius. Assuming that this estimate also works at
$\phi \neq 0$ and expanding $e^{n S_B} - e^{n S_0}$
to the first order in $B,$ one gets $ \delta
\rho_{xx}/ \rho
\sim -l/R_c = - \omega_c \tau.$

In fact, the physical picture of the phenomenon is more subtle.
The contribution of any trajectory with $\phi \neq 0$ is cancelled
in the first order in $B $ by the contribution of the
time-reversed trajectory, since the values of $ S_B - S_0$ are
opposite for these paths [ Fig1.~(d),(e)].  The cancellation does
not occur only at very small $\phi \sim \beta.$ The integration in
Eq. ~\ref{estimateB} over $\phi < \beta$  yields $\delta \rho_{xx}
/\rho \sim - \beta^2/\beta_0 $. Larger values of $\phi$ also give
a quadratic in $\beta$ contribution to the MR. This contribution
is positive and comes from the second order term in the expansion
of $e^ {n S_B}- e^{nS_0}$ in $B$.  It follows from our results
[Eqs.~\eqref{1},\eqref{f(z)}] that the contribution of small
angles is dominant resulting in a negative parabolic MR
\cite{lyap}.  We find that the parabolic MR crosses over to linear
at very small $\beta \approx 0.05 \beta_0, $ which explains why
the parabolic MR was not seen in numerical simulations
\cite{dmit1}   and experiment \cite{dot1}.

Next we sketch our calculations. We consider the
Lorentz gas at $T=0$, assuming that \mbox {$\beta \ll
1, \quad \beta_0 \ll 1.$ } In this case \cite{off},
$\delta \rho_{xx}/\rho = - (D-D_0)/D_0,$ where $D_0 =
v_F l_\text{tr}/2$ is the Drude diffusion coefficient
for $B=0,$ $l_\text{tr}=3l/4=3/8na$ is the momentum
relaxation length and $D$ is given by
\begin{equation}
D=\frac{1}{2}\int_0^\infty d t \langle \mathbf v(0)
\mathbf
v(t)\rangle= \frac{1}{2} \int d\mathbf{r} d\mathbf{v}
\langle G \rangle
\mathbf v \mathbf v_0.  \label{D} \end{equation}
Here $G=G(\mathbf v, \mathbf v_0,\mathbf{r})$ is the Fourier
transform (at $\omega=0$) of the retarded Green's function of the
Liouville equation and $\langle\dots\rangle$ stands for the
averaging over the positions of the disks.  The equation for $G$
reads
\begin{equation} \left( \hat L_0- \hat
T_{-}-\hat T_{+} \right) G =
\delta(\mathbf{r})\delta(\mathbf{v} -\mathbf{v}_0),
\label{Lio} \end{equation}
where $\hat L_0= \mathbf{v} \partial /\partial \mathbf{r} -
\omega_c [\mathbf v \times \partial /\partial \mathbf v]$ is the
Liouville operator of the free motion in the magnetic field.  The
interaction with disks is written in Eq. ~\ref{Lio} in the form of
a collision integral \cite{wei,mori}.  The scattering operators
$\hat T^{\pm}$ transform arbitrary function $f(\mathbf r, \mathbf
v)$ as follows,
\begin{eqnarray} \hat T^+ f(\mathbf{r},\mathbf{v})=
p_F \int d\mathbf{v^{\prime}} \sigma(\varphi)
\delta(\epsilon- \epsilon^{\prime})
n^+ f(\mathbf{r}, \mathbf{v^{\prime}}), \nonumber \\
\hat T^- f(\mathbf{r},\mathbf{v})= - p_F \int d\mathbf{v^{\prime}}
\sigma(\varphi) \delta(\epsilon- \epsilon^{\prime})
n^- f(\mathbf{r}, \mathbf{v} ), \label{+,-}
\end{eqnarray}
where $ n^{\pm} = \sum_i \delta(\mathbf{r}- \mathbf
R_i \pm \mathbf{a} ) .$
\begin{figure}
\includegraphics[width=0.17
\textwidth]{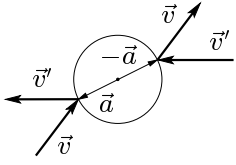}
\caption{Scattering of a particle on a hard disk. }
\label{fig2}
\end{figure}
Here $\mathbf R_i$ are the positions of the disks, delta-function
$\delta (\epsilon -\epsilon^{\prime})$ provides the energy
conservation, $p_F$ is the Fermi momentum, $\sigma(\varphi) =
(a/2) |\sin (\varphi/2)|$ is the differential cross-section of one
disk and $\varphi$ is the angle between $\mathbf{v^{\prime}}$ and
$\mathbf{v}$. The vector $ \mathbf a= \mathbf a(\mathbf
v^{\prime}, \mathbf v) = a(\mathbf v^{\prime} - \mathbf
v)/\sqrt{2(v^2- \mathbf v^{\prime} \mathbf v)}$ is pointing from
the center of a disk to the scattering point at the disk surface
[Fig.~2]. Physically, operator $\hat T^{+}$ describes influx of
particles  with velocity ${\bf v}$ at the point ${\bf R}_i -
\bf{a}$, while operator $\hat T^{-}$ describes the outflux of
particles  with velocity $ \bf{ v} $ at the point $\mathbf R_i +
\mathbf{a}.$ The Boltzmann equation is obtained from Eq.
~\ref{Lio} by averaging the Liouville operator over the positions
of the disks, yielding $ \langle \hat T^+\rangle
f(\mathbf{r},\mathbf{v})= p_F n \int d\mathbf{v^{\prime}}
\sigma(\mathbf{v}, \mathbf{v'}) \delta(\epsilon-
\epsilon^{\prime}) f(\mathbf r, \mathbf v^{\prime} ), \quad
\langle \hat T^-\rangle = - 1/\tau.$ Here $n= \langle n^{\pm }
\rangle $  is the concentration of the disks and $1/\tau = n v 2
a$ is the inverse full scattering time.
\begin{figure}
\includegraphics[width=
0.4
\textwidth]{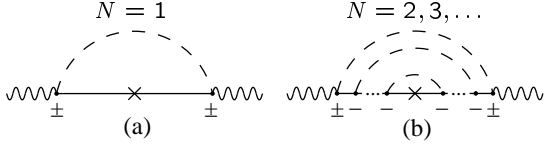}
\caption{Diagrams, corresponding to the process shown in
Fig.~1(a).  Diagram (a) does not take into account effect of
``empty corridor'' and should be
renormalized by (b). }
\label{fig3} \end{figure}
Introducing now $\delta \hat T^{\pm} = \hat T^{\pm} - \langle \hat
T^{\pm} \rangle$ and writing a formal solution of \eqref{Lio}, $
\hat G =(\delta +\hat L_0 -\hat T^- -\hat T^+)^{-1}$ as a series
in $\delta \hat T^{\pm},$ we get
\begin{equation} \langle \hat G \rangle= \hat G_0
+\sum_{\alpha,\gamma = \pm} \hat G_0 \langle \delta
\hat T^{\alpha}\hat  G_0
\delta \hat T^{\gamma} \rangle \hat G_0 + \cdots,
\label{series}
\end{equation}
where $\hat G_0=(\delta +\hat L_0 + 1/\tau -\langle \hat
T^{+}\rangle )^{-1} $ is the Green's function of the Boltzmann
operator (here $\delta \to 0$).  Eq. ~\ref{series} gives a regular
way to calculate correlations, which are absent in the Boltzmann
picture.

Consider first the case $B=0.$ Substituting the first term in the
right hand side of Eq. ~\ref{series} into Eq. ~\ref{D}, we get
$D=D_0.$ The second term in Eq. ~\ref{series} describes the
memory-effect due to diffusive returns. As discussed above, the
main contribution comes from returns after a single scattering.
This process is described by the diagram Fig.~3(a). The dashed
line corresponds to the pairings $\langle \delta \hat T^{\alpha}
\delta \hat T^{\gamma}\rangle $ ($\alpha,\gamma =\pm$), external
wavy lines to the diffusion propagators $\hat G_0.$ The internal
line corresponds to the Boltzmann propagator truncated at one
scattering $\hat G_- \langle \hat T^+ \rangle \hat G_-,$ where
$\hat G_-= (L_0 +1/\tau)^{-1}$ is the ballistic propagator and
$\langle \hat T^+ \rangle $ stands for one scattering event ($G_-$
are shown by solid lines and $\langle \hat T^+ \rangle $ by the
cross). This diagram yields $ \delta \rho/\rho = -\delta D /D =
(2\beta_0/3)\ln(1/2\beta_0)$ \cite{dorfman,wei,wei1,hauge,peierls}
. The terms of the $N- \text{th}$ order in Eq. ~\ref{series}
contain $N$ pairings ($N$ dashed lines) and are typically small as
$\beta_0^N.$ However, there is a series of diagrams, shown in
Fig.~3(b), whose contribution is of the order $\beta_0$
\cite{wei1}. The internal dashed lines in this series only contain
pairings $\langle \delta \hat T^- \delta \hat T^- \rangle.$
Summing the diagrams Fig.~3(b) together with Fig.~3(a), one gets
an exact equation
\begin{align} &\frac{ \delta \rho}{\rho}= \frac {n
l_\text{tr}}{4l} \text{Re} \int_a^{\infty}
\frac{dr}{r}e^{-2r/l}
\int_0^{2\pi} d \varphi_0 \int_0^{2\pi} d \varphi_f
\sigma(\varphi_0) \sigma(\varphi_f)
\nonumber \\ &(1- e^{i \varphi_0}) (1- e^{i
\varphi_f}) e^{ nS_0 (r, \phi_0
+\phi_f)} = \frac{2\beta_0}{3}\ln \left
(\frac{C}{2\beta_0 }\right), \label{fin1} \end{align}
instead of qualitative estimate Eq. ~\ref{estimate}.
\begin{figure}
\includegraphics[width=0.45\textwidth]{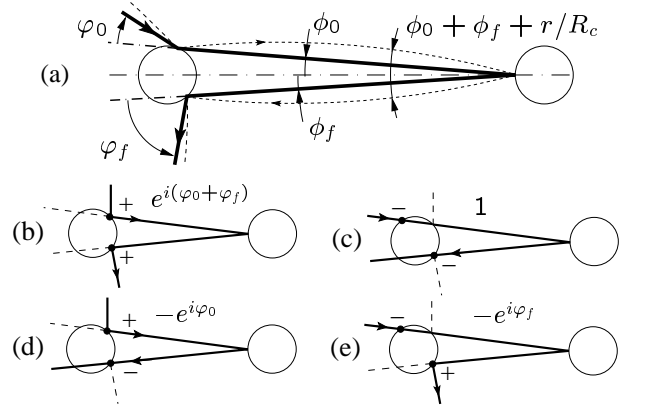}
\caption{ Backscattering process is parameterized by the angles
$\varphi_0,\varphi_f.$ The magnetic field changes the
backscattering angle $\phi=\phi_0 +\phi_f +r/R_c.$ The solid
(dashed) line in (a) represents electron trajectory for $B=0$
($B\neq 0$).  Different processes contributing to MR are shown in
(b)-(e).} \label{fig4}
\end{figure}
Here $\varphi_0 , \varphi_f$ are the scattering angles
[Fig.~4(a)], $ \phi_0 \approx (a/r) \cos(\varphi_0/2),$   \mbox{ $
\phi_f \approx(a/r) \cos(\varphi_f/2) $} and $C \approx 1.8$ Thus,
addition of the series Fig.~3(b) to Fig.~3(a) leads to the
following renormalization: $ \ln(1/2\beta_0) \to \ln(C/2\beta_0).$
Physically, the series Fig.~3(b) accounts for the effect of the
``empty corridor''. The $N-$th order term in this series
corresponds to $N-1$ term in the Taylor expansion of the $\exp(n
S_0)$ in Eq. ~\ref{fin1}.  Four terms in the product $(1-e^{i
\varphi_0})(1-e^{i \varphi_f})= 1 -e^{i\varphi_0}-e^{i\varphi_f}+
e^{i(\varphi_0 +\varphi_f)}$ correspond to four combinations of
$(\pm,\pm)$ at the ends of external dashed lines in the diagrams
shown in Fig.~3. They are connected with four different types of
correlation at a given point $\mathbf r.$ The diagram $(+,+) $
[Fig.~4(b)] corresponds to the process, where an electron has two
real scatterings on a disk placed at point $ \mathbf r .$ The
diagram $(-,-)$ [Fig.~4(c)] does not correspond to any real
scattering at point $\mathbf r.$ It just allows us to calculate
correctly the probability for an electron to pass twice the region
of the size $a$ around point $\mathbf r$ without scattering.  To
interpret the diagram $(+,-),$ note that in the Boltzmann picture,
which neglects correlations, the following process is allowed.  An
electron scatters on a disk and later on passes through the region
occupied by this disk without a scattering [Fig.~4(d)]   The
diagrams $(+,-)$  correct the Boltzmann result by substracting the
contribution of such unphysical process. Analogous consideration
is valid for diagram $(-,+)$ shown in Fig.~4[e].

For $B \neq 0$ the sum of diagrams shown in Fig.~3 can be
expressed as an integral over angles $\varphi_0, \varphi_f$
(scattering angles for $B=0$). The only difference from Eq.
~\ref{fin1} is that one should replace $S_0 \to S_B.$ For $\beta
\ll1$ the overlap area can be calculated as $ S_B (r, \phi) =
\int_0 ^r d x h(x) , $ where $h(x) \approx \left(2a - \left\vert
\phi x - x^2/R_c \right\vert\right) \theta \left(2a - \left\vert
\phi x - x^2/R_c \right\vert \right)$, $\theta$ is the Heaviside
step function and $ \phi =\phi_0 + \phi_f +r/R_c$ [Fig.~4(a)]. The
value of $\delta \rho_{xx}/\rho$ is obtained  from Eq. ~\ref{fin1}
by replacing $e^{n S_0}$ to $e^ {n S_B} - e^{n S_0}.$ Introducing
dimensionless variables $T=r/l, \ z= \beta/\beta_0$ we get Eq.
~\ref{1}, where function $f(z)$ is given by
\begin{align}
&f(z)= \frac{3}{32} \int_0^{\infty}\frac{dT}{T}
e^{-2T} \int_0^{2\pi} d \varphi_0 \int_0^{2\pi} d
\varphi_f
\nonumber \\ & \cos\left(\frac{\varphi_0
+\varphi_f}{2}\right)
\sin^2\left(\frac{\varphi_0}{2}\right) \sin^2
\left(\frac{\varphi_f}{2}\right) \left( e^{ s_z} -
e^{ s_0}\right).  \label{f(z)} \end{align}
Here
\begin{align} &s_z= \int_0^{T} dt \left(1-\left\vert
\zeta t -\frac{z t^2}{2} \right\vert \right) \theta
\left(1- \left\vert \zeta t -\frac{z t^2}{2}
\right\vert \right),
\nonumber \\ &\zeta=\frac{\cos(\varphi_0/2)
+\cos(\varphi_f/2)}{2T} +\frac{z T}{2}, \quad s_0=
s_{z \to 0}.  \label{s_z} \end{align}
Function $f(z)$ has the following asymptotics
\begin{equation} f(z)= \left \{ \begin{array}{ll}
0.33 z^2 & \text{for $z \lesssim 0.05$} \\ 0.032\ (z
-0.04) &\text{for $ 0.05 \lesssim z \lesssim 2$} \\
0.39 - 1.3 /\sqrt{z} & \text{for $ z \to \infty$}.
\end{array} \right.
\label{asimpt} \end{equation}
Note that there is a parametrically small nonanomalous correction
to Eq. ~\ref{1} due to returns after multiple scatterings, $\delta
\rho_{xx}^{\prime}/\rho \approx - 0.2 \beta_0 \beta^2$
\cite{dmit1}.  To compare the results of simulations \cite{dmit1}
with the theoretical results in a wider region of parameters
$\beta, \beta_0,$ we substract $\delta \rho_{xx}^{\prime}/\rho$
from the numerical curves.  Theoretical and numerical \cite{dmit1}
results  are plotted in Fig.~5. in the universal units, $\delta
\rho_{xx}/ \rho \beta_0  $ versus $z=\beta/\beta_0.$ It is seen,
that the theoretical and numerical results are in a very good
agreement.
\begin{figure}
\includegraphics[width=0.4135\textwidth]{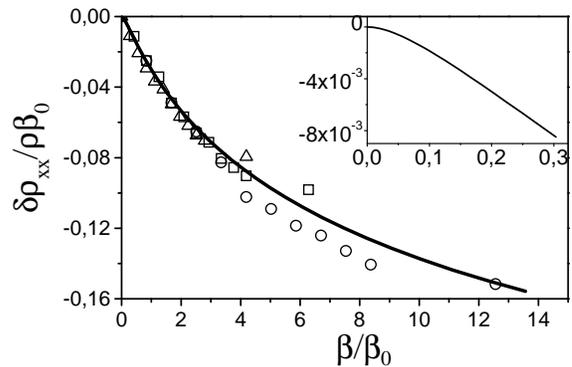}
\caption{The value of $\delta\rho_{xx}/ \rho \beta_0 $ from
Eqs.~\eqref{1},~\eqref{f(z)} (solid line) shown as a function of
$\beta/\beta_0$ together with the results of numerical simulations
\cite{dmit1} presented for different values of $\beta_0$
(triangles for $\beta_0= 0.09$ , boxes for $\beta_0= 0.06$ ,
circles for $\beta_0=0.03$). Data for all numerical curves are
shown for $\beta < 0.3.$ Inset: The crossover from quadratic to a
linear dependence at $\beta/\beta_0 \sim 0.05.$ This crossover was
not resolved in numerical simulations. } \label{fig5}
\end{figure}
The comparison with the experiment \cite{dot1} is
more difficult, because of the 50~\% uncertainty in
the sizes of the antidots.  However, a good agreement
with the experiment can be achieved by appropriate
choice of $a$ in the uncertainty interval
\cite{dmit1}.

Note finally that we fully neglected quantum effects.
This is possible when $a \gg \sqrt{\lambda_F l}$
($\lambda_F$ is a Fermi wavelength). This criterion
ensures that diffraction effects on the edges of the
disks are not relevant at the scales of the order of
$l.$ In the opposite case, $a \ll \sqrt{\lambda_F
l},$ the diffraction should destroy the ``corridor
effect'', does suppressing the anomalous MR. The
detailed analysis of quantum effects will be
presented elsewhere.

In summary, we have proposed a theory of the negative
anomalous MR in the Lorenz gas. The analytical
expression for the MR
[Eqs.~\eqref{1},~\eqref{f(z)},~\eqref{asimpt}] has
been derived.

We thank M.I.~Dyakonov for insightful discussions and R.~Jullien
for providing us with the numerical data. We are also grateful to
I.V.~Gornyi and D.G.~Polyakov for useful comments. The work was
partially supported by RFBR and INTAS.


\begin{thebibliography}{99}
\bibitem{lyap} For very small $\beta$ contribution of
Lyaponov trajectories becomes important.  This contribution is
parametrically small compared to Eq.\eqref{1} for $\beta \gtrsim
\beta_0 ^2.$
\bibitem{off} For $\beta \ll 1$ one can neglect the
off-diagonal part of the conductivity tensor.
\end{thebibliography}
\end{document}